\documentclass[runningheads]{llncs}
\usepackage{amsmath,amsfonts,amssymb}
\usepackage{bm}
\usepackage{cite}
\usepackage{diagbox}
\usepackage{doi}
\usepackage[english]{babel}
\usepackage{float}
\usepackage{footnote}
\usepackage{graphicx}
\usepackage{marvosym}
\usepackage{multirow}
\usepackage{textcomp}
\usepackage{xcolor}
\newtheorem{Def}{Definition}

\newtheorem{Theo}{Theorem}
\newenvironment{proofsketch}{
	\proof}{\endproof}
\newcommand{\tabincell}[2]{ \begin{tabular}{@{}#1@{}}#2\end{tabular} }
\begin{document}
\title{Two-Server Verifiable Homomorphic Secret Sharing for High-Degree Polynomials}
\author{Xin Chen\inst{1,2,3} \and
	Liang Feng Zhang\inst{1}$^{(\textrm{\Letter})}$}
\institute{
School of Information Science and Technology, ShanghaiTech University, Shanghai, China \\
\email{\{chenxin3,zhanglf\}@shanghaitech.edu.cn}\\
\and
Shanghai Institute of Microsystem and Information Technology, Chinese Academy of Sciences, Shanghai, China
\and
University of Chinese Academy of Sciences, Beijing, China
}
\titlerunning{2SVHSS for High-Degree Polynomials}
\maketitle
\begin{abstract}
Homomorphic secret sharing (HSS) allows multiple input clients to secret-share their data among multiple servers such that  each server is able to locally compute a function on its  shares to obtain a partial result  and all partial results enable the reconstruction of the function's value on
the outsourced data by an output client.
The existing HSS schemes for {\em high-degree} polynomials
either {\em require a large number of servers} or {\em lack verifiability}, which is essential for
ensuring the correctness of the outsourced computations.
In this paper, we propose  a two-server verifiable HSS (VHSS)  model and construct a scheme that   supports the computation of high-degree polynomials.
The degree of the outsourced polynomials can be as high as a polynomial in the system's security parameter. Despite of using only 2 servers, our VHSS ensures that each single server learns no information about the outsourced data and no single server is able to persuade the client to output a wrong function value.
Our VHSS is significantly more efficient. When computing degree-7 polynomials, our scheme could be 3-10 times faster than the previously  best construction.

\end{abstract}

\section{Introduction}

 In the context of outsourcing computations, the outsourced data   may be leaked  and the servers may be hijacked or return incorrect results for   economic reasons (such as  saving the computing resources). How to ensure the \textit{privacy} of the outsourced data and the {\em integrity} of the outsourced computations  are two   top security issues.

A general  method of protecting the
 privacy of the data in outsourcing computation  is by using
  the homomorphic encryption (HE), which allows the cloud
   servers to compute  a function $f$ on the ciphertexts ${\sf Enc}(x_{1})$, $\dots$, ${\sf Enc}(x_{n})$ to get a ciphertext of the function value $y=f(x_{1},\dots,x_{n})$. The early HE schemes   \cite{GM84,P99}  only support degree-1 computations on the encrypted data. Gentry \cite{G09} proposed the first fully homomorphic encryption (FHE) scheme that allows the  computation of any boolean circuits on encrypted data. Although the efficiency of FHE has been significantly improved
 in  \cite{BV14,BGV12} during the past years, FHE is still
impractical from the performance perspective\cite{MSM17}.
As a multi-server counterpart of HE that is more efficient, the HSS of Boyle et al.\cite{BGI16}  allows the input clients to secret-share their data among multiple servers such that
 upon request each server is able to compute a partial result and
 all partial results suffice to   reconstruct the correct function value by
 an output client.
  Most of the existing HSS schemes are designed for computing
   specific functions, such as the affine functions\cite{B86}, the point functions\cite{BGI15}, the selection functions\cite{CM99,E09}, and the depth-2 boolean circuits \cite{BIK12}.
Recent  works \cite{LMS18,CF15,BFC17,CZ19,BKS19} have  focused on the construction of HSS schemes that support high-degree polynomial computations.

While HE and HSS provide   easy solutions to the data privacy problem in
outsourcing  of computations,  they cannot help the output client
verify whether the servers have done their computations correctly.
 Tsaloli et al.\cite{TLM18} proposed the notion of verifiable HSS (VHSS)   which
 additionally add verifiability of the servers' results to the HSS schemes.
 Unfortunately, their construction only supports the product computations  over a
  multiplicative Abelian group and has been proved to be insecure \cite{HZ20}.
   Yoshida and Obana \cite{YO18} constructed verifiably  multiplicative secret sharing schemes  that enable the computation of polynomials over the shared data.
   However, the degrees of their polynomials  cannot exceed  the number of servers.

Therefore, if we restrict our attention to HSS schemes for high-degree polynomial computations, the state of the art offers either the construction \cite{BKS19}
that has no verifiability or the constructions \cite{YO18} that require a very large number of servers. In this paper, we shall focus on the construction of {\em two-server}
{\em verifiable} HSS schemes that allow {\em high-degree} polynomial computations.

\subsection{Our Contributions}
In this paper, we propose  a  two-server verifiable homomorphic secret sharing (2SVHSS) model. Our model involves three kinds of parties: a set of {\em input clients}, a set of {\em servers}, and an {\em output client}. Each input client   uses a public key to encrypt its data as shares  to the servers; upon the request of computing a  function
on the outsourced  data, each server performs a computation  on its own shares   and produces a partial result; finally, the output client reconstructs the function value from all partial results.
 A 2SVHSS scheme in our model should satisfy the properties of \textit{correctness}, \textit{semantic security}, \textit{verifiability}, \textit{context hiding} and \textit{compactness}. The correctness property requires that whenever the scheme
 is correctly executed, the output client will always reconstruct
 the correct function value. The semantic security requires that each server learns no information about the outsourced data. The verifiability requires that no malicious server is able to persuade the output client to reconstruct a wrong function value. The context hiding property requires that the output client  learns no more information about the outsourced data than what is implied by the function value. This property is specifically interesting when the output client is not one of the input clients. The compactness property requires  that  the output client's workload in a protocol execution should be substantially less than that required by the native computation of the function. This property is essential for HSS's applications in outsourcing computation.
In the proposed model we  construct  a 2SVHSS scheme that allows the
computation of  polynomials over the   outsourced data.
The degrees of these polynomials can be as {\em high} as a polynomial in the system's security parameter.
Achieving {\em verifiability} in two-server HSS for {\em high-degree} polynomials allows us to distinguish between this work and the existing ones.

\subsection{Our Techniques}
Our construction is based on the HSS of \cite{BKS19}. The core technique of the HSS scheme of \cite{BKS19} is a public-key encryption scheme supporting nearly linear decryption (PKE-NLD). Let $R=\mathbb{Z}[X]/(x^N+1)$, where $p,q\in\mathbb{N}$, $p|q$ and $1\ll p\ll q$. Informally, a public-key encryption scheme supports nearly linear decryption for a message $x\in R_p$ if the secret key is ${\bf{s}}=(s_1,s_2)=(1,s) \in R_p^2$, and for any ciphertext ${\bf{c}}\in R_q^2$ encrypting $x$, $\langle {\bf{s},\bf{c}} \rangle=(q/p)\cdot x +e \mod q$ for some ``small" noise $e\in R$. The input clients use PKE-NLD and the public key $\sf pk$ to encrypt $x\cdot{\bf{s}}$ as ${\bf C}^x=({\bf c}^{x\cdot 1},{\bf c}^{x\cdot s})\in  R_q^{2\times 2}$, without knowing the secret key ${\bf s}$. In order to implement homomorphic computation on the ciphertext, they randomly split the secret key $\bf s$ into a pair of evaluation keys $({\sf ek}_1,{\sf ek}_2)\in R_q^{2\times 2}$ such that ${\bf s}=({\sf ek}_1+{\sf ek}_2)\mod q$. For every $b\in \{1,2\}$, the server $b$ uses ${\sf ek}_b$ to compute a share ${\bf t}_b^{x}$ of $x\cdot\bf{s}$, and returns it to the output client. It is clear that for $i\in\{1,2\}$, $\langle {\bf t}_0^{x},{\bf c}^{x'\cdot s_i} \rangle+\langle {\bf t}_1^{x},{\bf c}^{x'\cdot s_i} \rangle=\langle x\cdot {\bf s}, {\bf c}^{x'\cdot s_i}\rangle \approx (q/p)\cdot xx'\cdot s_i$ over $R_q$.

The HSS of \cite{BKS19} supports the computation of restricted multiplication straight-line programs\cite{BGI16,CR91}. To enable the computation of polynomials, we added homomorphic multiplication by known constants to \cite{BKS19} as follows. For every $b\in \{1,2\}$, the server $b$ computes the secret share ${\bf t}_b^{c\cdot x}$ (in fact, ${\bf t}_b^{c\cdot x}=c\cdot{\bf t}_b^x$) of $c \cdot x\cdot\bf{s}$ (where $c\in R_p$) by using $c\cdot {\sf ek}_b$. The correctness of the computation can be established as follows: ${\bf t}_1^{c\cdot x}+{\bf t}_2^{c\cdot x}=c\cdot({\bf t}_1^x+{\bf t}_2^x)=(cx\cdot 1,cx\cdot s)$. To achieve verifiability, we add a verification key and a pair of additional evaluation keys.
More precisely, the verification key ${\sf vk}=(\hat{s},\hat{s}\cdot s)$ is obtained by multiplying the secret key ${\bf s}=(1,s)$ with a randomly chosen value $\hat{s}\in R_p$; and the additional evaluations keys $(\hat{{\sf ek}_1},\hat{{\sf ek}_2})$ are additive shares of ${\sf vk}$, i.e., ${\sf vk}=(\hat{{\sf ek}_1}+\hat{{\sf ek}_2}) \mod q$. With the additional evaluation keys, the servers can manage to compute $\tau=\hat{s}\cdot f(x_{1},\dots,x_{n})$. The output client can decide whether $f(x_{1},\dots,x_{n})$ is computed correctly by verifying the equation $\tau=\hat{s}\cdot f$. A successful attack of the verifiability by a malicious server requires that server to guess $\hat{s}$ correctly. However, that will happen with at most a negligible probability, because $\hat{s}$ is randomly split into two additive shares and sent to two servers respectively, a single server cannot obtain effective information about $\hat{s}$. Our verification process only needs to perform a fixed number of additions and multiplications on $R$ and does not need to perform decryption (including modular exponentiation) like \cite{LMS18,CF15,BFC17}. Therefore, our verification process consumes few resources which is very friendly to the output clients with limited computing power.
\subsection{Applications}
Our scheme has interesting applications in many scenarios such as secure computation \cite{BCG17,BGI16}, private manipulation of remote databases\cite{CBM15}, and generating correlated random variables \cite{BCG17}. In addition, it can be used in industry.

\hspace{-0.52cm}{\bf Deep Neural Networks:}  The scheme of \cite{ZCL20} has been using uses FHE to enable the evaluation of deep neural networks over the private data.  Our scheme can provide an efficient alternative to FHE, and our scheme is verifiable and ensures the users to get the correct results from untrusted servers.

\hspace{-0.52cm}{\bf Smart Grid Network:} Smart grid networks are envisioned to be the next generation power supply networks. The smart grid deploys sensors on the consumers to collect real-time data which will be uploaded to the servers. The control center requires the servers to perform some computations on the collected data for further analysis, such as regulating the supply of power. Khurana et al. \cite{KHL10} pointed out that the data stored on the servers will reveal their private information. A more serious problem is that the servers may be hijacked which will cause the control center to make a wrong judgment and cause catastrophic consequences. Such security problems can be solved by our scheme, which can preserve the privacy of the consumers by encrypting the collected data and allow the control center to verify the result.

Our scheme can be used to provide both data privacy, results verification and efficient computation in many other systems such as the healthcare systems\cite{SCS18} and the industrial control systems\cite{SWW15}.

\subsection{Related Work}
Recent works\cite{CF15,BFC17,LMS18} are to transform the level-$k$ HE scheme to support the computation of high-degree polynomials. Informally, if an HE scheme can support computations of degree $\leq k$, we call it a level-$k$ HE scheme. Although \cite{CF15,BFC17,LMS18} achieve the purpose of computing polynomials through level-$k$ HE, their constructions do not support verification and are only effective for low-degree polynomials. Some works\cite{YO18,CZ19} constructed multi-server schemes with verifiability. However, the degree of their polynomials can not exceed the number of servers.

\begin{table}[H]
	\centering
	\caption{Comparisons with Existing HSS Schemes}
	\begin{tabular}{cccccccc}
		\hline
		& $n$ & $m$   & $d$             & {\tabincell{c}{semantic\\ security}}      & {verifiability} & {\tabincell{c}{context\\ hiding}} & {compactness}   \\ \hline
		\cite{CF15} & *   & 2       & $2k$            & $\checkmark$ & $\times$      & $\checkmark$   & $\checkmark$       \\
		\cite{BFC17}     & 1   & 1       & $2k$            & $\checkmark$ & $\times$      & $\checkmark$   & $\times$          \\
		\cite{LMS18}         & *   & $m$     & $(k+1)m-1$      & $\checkmark$ & $\times$      & $\checkmark$   & $\checkmark$      \\
		\cite{YO18}  & *   & $m$  & $<m$          & $\checkmark$ & $\checkmark$  & $\checkmark$   & $\checkmark$   \\
		\cite{CZ19}     & 1   & $m$    & $m$             & $\checkmark$ & $\checkmark$  & $\checkmark$   & $\times$            \\
		Ours            & *   & 2    & ${\sf poly}(\lambda)$ & $\checkmark$ & $\checkmark$  & $\checkmark$   & $\checkmark$   \\ \hline
	\end{tabular}
\end{table}
Table 1 shows the comparisons between our scheme and some existing representative schemes for $n$ input clients, $m$ servers, and degree-$d$ polynomials, where ``$n=*$'' means that the number of allowed input clients is unbounded. Compared with these representative works, our scheme satisfies all properties in comparison and supports polynomials of highest degree.

\subsection{Organization}
In Section 2 we introduce the techniques that will be used in our construction. In Section 3 we formally define 2SVHSS. In Section 4 we give both construction and analysis of our 2SVHSS. In Section 5 we implement our scheme and compare with the best existing schemes in terms of efficiency. Finally, Section 6 contains our concluding remarks.

\section{Preliminaries}
We denote with $\lambda \in \mathbb{N}$ a security parameter, and use ${\sf poly}(\lambda)$ to denote any function bounded by a {\em polynomial} in $\lambda$. We say that a function is {\em negligible} in $\lambda$ if it vanishes faster than the inverse of any polynomial in $\lambda$, and denote a negligible function by $\sf negl$.  We use PPT for probabilistic polynomial-time. For a positive integer $n$, we denote by $[n]$ the set $\{1,2,\dots,n\}$.
For a real number $x\in\mathbb{R}$, by $\lfloor x \rceil \in \mathbb{Z}$ we denote the integer closest to $x$. Let $R=\mathbb{Z}[X]/(X^N+1)$, where $N\in\mathbb{N}$ with $N\leq {\sf poly}(\lambda)$ is a power of 2. For $x\in R$ with coefficients $x_1,\dots,x_N$, the {\em infinity norm} of $x$ is defined as $\|x\|_{\infty}:={\rm{max}}_{i=1}^N|x_i|$.
For $p\in\mathbb{N}$, by $R_p$ we denote $R/pR$. We agree that the coefficients of any element in $R_p$ are in the interval $(-\lfloor p/2 \rceil,\dots,\lfloor (p-1)/2 \rceil]$. We define the {\em size} of a polynomial as the total number of multiplication and addition operations it contains. The size of the polynomial $f$ is denoted as ${\sf size}(f)$.
Let $X,Y$ be two probability distributions over the same sample space $U$. We define the {\em statistical distance} between $X$ and $Y$ as
${\sf SD}[X,Y]:=\frac{1}{2}\sum_{u\in U}|\Pr[X=u]-\Pr[Y=u]|.$

\subsection{Public-Key Encryption with Nearly Linear Decryption}

Boyle et al. \cite{BKS19} gave an instantiation of PKE-NLD that satisfies the following properties: First, it allows anyone to encrypt certain key-dependent messages without the secret key. Second, it allows distributed decryption of ciphertexts.

The PKE-NLD of \cite{BKS19} was instantiated with Ring-LWE\cite{BV11,LPR13} and parametriz-ed by modulus values $p,q \in \mathbb{N}$, and bounds $B_{\sf sk},B_{\sf ct} \in \mathbb{N}$, where $p|q$, $p\geq \lambda^{\omega(1)}$, $q/p\geq \lambda^{\omega(1)}$ and $B_{\sf sk},B_{\sf ct} \leq {\sf poly}(\lambda)$, as well as a ring $R=\mathbb{Z}[X]/(X^N+1)$, where $N$ is a power of 2. The secret key $\bf s$ satisfies $\| {\bf s}\|_{\infty} \leq {B}_{\sf sk}$. The error $e$ satisfies $\|{e}\|_{\infty} \leq {B}_{\sf ct}$.

In the PKE-NLD instantiation of \cite{BKS19}, $\mathcal{D}_{\sf sk}$ is a secret-key distribution such that each coefficient of the secret key $s$ is uniformly distributed over $\{0,\pm1\}$, subject to the constraint that $h_{\sf sk}$ out of the coefficients of $s$ are non-zero. $\mathcal{D}_{\sf err}$ is an error distribution where each coefficient is a rounded Gaussian with a parameter $\sigma$,  which gives $B_{\sf err}=8\sigma$ as a high-probability bound on the $l_\infty$ norm of samples from $\mathcal{D}_{\sf sk}$, with failure probability about $2^{-49}$.  The instantiation of PKE-NLD can be described as follows:
\begin{itemize}
	\item ${{\sf PKE.Gen}(1^\lambda):}$ Sample $a\leftarrow R_q$, ${s}\leftarrow \mathcal{D}_{\sf sk}$, $e\leftarrow \mathcal{D}_{\sf err}$ and compute $b=a\cdot{s}+e$ in $R_q$. Let ${\bf s}=(1,{s})$ and output ${\sf pk}=(a,b)$, ${\sf sk}={\bf s}$.
   		
	\item ${{\sf PKE.Enc}({\sf pk},m):}$ To encrypt $m\in R_p$, sample $v\leftarrow \mathcal{D}_{\sf sk}$, $e_0,e_1\leftarrow \mathcal{D}_{\sf err}$. Output the ciphertext $(c_0,c_1)\in R_q^2$, where $c_1=-av+e_0$ and $c_0=bv+e_1+(q/p)\cdot m$.

	\item ${{\sf PKE}.{\bf OKDM}({\sf pk},m):}$ Compute ${\bf c}^0={\sf PKE.Enc}(0)$ and ${\bf c}^m={\sf PKE.Enc}(m)$. Output the tuple $({\bf c}^m,{\bf c}^0+(0,(q/p)\cdot m))$ as an encryption of $m\cdot {\bf s}$.

	\item ${{\sf PKE.DDec}(b,{\bf t}_b,{\bf c}^x):}$ Given $b\in[2]$, a ciphertext ${\bf c}^x=(c_0,c_1)$ and a share ${\bf t}_b=(t_{b,0},t_{b,1})$ of $m\cdot{\bf s}$. Output $d_b=(\lfloor (p/q)\cdot(c_0\cdot t_{b,0}+c_1\cdot t_{b,1})\rceil \mod p) \mod q$.
\end{itemize}

Without accessing the secret key, anyone can compute the encryption of any linear function of the secret key through key-dependent message (KDM) oracle.

\hspace{-0.52cm}{\bf Nearly linear decryption to the message} $x\cdot s_j:$ for any $\lambda \in \mathbb{N}$, for any $({\sf pk},{\bf s})\leftarrow{\sf PKE.Gen}(1^\lambda)$, and for any ${\bf c}_j\leftarrow{\sf PKE.OKDM}({\sf pk},x,j)$, it holds
	$\langle{\bf s},{\bf c}_j\rangle=(q/p)\cdot(x\cdot s_j)+e \mod q$
	for some $e\in R$ with $\|e\|_\infty \leq B_{\sf ct}$, where $j\in[2]$.
	
	\hspace{-0.52cm}{\bf Security}: for any $\lambda \in \mathbb{N}$ and any PPT adversary $\mathcal{A}$, ${\bf Adv}_{{\mathcal{A},\sf PKE.OKDM}}^{\sf kdm-ind}(\lambda):=\left| \Pr \left[ {\bf Exp}_{{\mathcal{A},\sf PKE.OKDM}}^{\sf kdm-ind}(\lambda)=1\right]-1/2\right|\leq {\sf negl}(\lambda)$, where ${\bf Exp}_{{\mathcal{A},\sf PKE.OKDM}}^{\sf kdm-ind}(\lambda)$ is defined as follows:
	
	\begin{center}
	\begin{tabular}{lll}
		${{\bf Exp}_{{\mathcal{A},\sf PKE.OKDM}}^{\sf kdm-ind}(\lambda):}$            & \quad\quad & ${\mathcal{O}_{\sf KDM}(x,j):}$     \\
		$({\sf pk},{\sf sk})\leftarrow {\sf PKE.Gen}(1^\lambda)$                                & \quad\quad & {If} $\beta=0$ {return} ${\sf PKE.OKDM}({\sf pk},x,j)$.                               \\
		$\beta\leftarrow\{0,1\}$                                                                & \quad\quad & {Else} {return} ${\sf PKE.Enc}({\sf pk},0)$.   \\
		$\beta'\leftarrow \mathcal{A}^{\mathcal{O}_{\sf KDM}(\cdot,\cdot)}(1^\lambda,{\sf pk})$ & \quad\quad &      \\
		{If} $\beta=\beta'$ {return} $1$.                                                        & \quad\quad & \quad     \\
		{Else return} $0$.                                                                       & \quad\quad &
	\end{tabular}
	\end{center}
	By ${\sf PKE}.{\bf OKDM}({\sf pk},x)$ we denote the KDM oracle that
	returns a componentwise encryption of $x\cdot {\bf s}$, i.e. the matrix $({\sf PKE.OKDM}({\sf pk},x,1)$, ${\sf PKE.OKDM}({\sf pk},x,2))$ $\in R_{q}^{2\times 2}$.

 The second property that PKE-NLD needs to satisfy is that it allows two non-communicating servers to perform decryption distributively.

\hspace{-0.52cm}{\bf Distributed decryption of sums of ciphertexts}:
Let $B_{\sf add} \in \mathbb{N}$ be a polynomial in $\lambda$. Then there exists a deterministic polynomial time decryption procedure $\sf PKE.DDec$ with the following properties: for all $x \in R_p$ with $p/\|x\|_\infty \geq \lambda^{\omega(1)}$ and $q/(p\cdot \|x\|_\infty)\geq \lambda^{\omega(1)}$, for all $({\sf pk},{\bf s})$, for all messages $m_1,\dots,m_{B_{\sf add}} \in R_p$, for all encryption ${\bf c}_i$ of $m_i$ that are either output of $\sf PKE.Enc$ or $\sf PKE.OKDM$ (in that case we have $m_i=x_i\cdot s_j$ for some $x_i\in R_p$ and $j\in[2]$), for the shares ${\bf t}_1,{\bf t}_2 \in R_q^2$ that were randomly chosen subject to ${\bf t}_1+{\bf t}_2=x\cdot{\bf s} \mod q$, for ${\bf c}:=\sum_{i=1}^{B_{\sf add}}{\bf c}_i$ and $m:=\sum_{i=1}^{B_{\sf add}}m_i$, ${\sf PKE.DDec}(1,{\bf t}_1,{\bf c})+{\sf PKE.DDec}(2,{\bf t}_2,{\bf c})=x\cdot m \mod q$ with probability at least
$$1-N\cdot(N\cdot B_{\sf add}\cdot\|x\|_{\infty}\cdot B_{\sf ct}\cdot p/q+\|x\cdot m\|_\infty/p+p/q+1/p)\geq 1-\lambda^{-\omega(1)}$$
over the randomly choice of the shares ${\bf t}_1,{\bf t}_2$.
For ${\bf C}=({\bf c}_1|{\bf c}_2)\in R_p^{2\times 2}$ by ${\bf m}\leftarrow {\sf PKE}.{\bf DDec}(b,{\bf t}_b,{\bf C})$, we denote the componentwise decryption ${\bf m}\leftarrow({\sf PKE.DDec}(b,{\bf t}_b,{\bf c}_1),{\sf PKE.DDec}(b,{\bf t}_b,{\bf c}_2))\in R_p^2.$

\section{Two-Server Verifiable Homomorphic Secret Sharing}
A two-server verifiable homomorphic secret sharing (2SVHSS) scheme involves three kinds of parties: multiple input clients, two non-communicating servers and an output client. In a 2SVHSS scheme, any client that gets the public key can encrypt input value as ciphertexts and share between the two servers. The output client asks servers to compute the result and uses the verification key to verify the result.
\begin{Def}[\bf Two-server verifiable homomorphic secret sharing]
	
A 2S-VHSS scheme ${\sf 2SVHSS}$ for a function family $\mathcal{F}$ over a ring $R$ with input space $\mathcal{I}\subseteq R$ consists of four PPT algorithms $\sf(2SVHSS.Gen$, $\sf 2SVHSS.Enc$, $\sf 2SVHSS.Eval$, $\sf 2SVHSS.Ver)$	with the following syntax:
	
	\begin{itemize}
		
		\item ${\sf 2SVHSS.Gen}(1^\lambda)$: On input a security parameter $1^\lambda$, the key generation algorithm outputs a public key $\sf pk$, a verification key $\sf vk$ and a pair of evaluation keys $({\sf ek}_1,{\sf ek}_2)$.
		
		\item ${\sf 2SVHSS.Enc}({\sf pk},x)$: On input a public key $\sf pk$ and an input value $x \in \mathcal{I}$, the encryption algorithm outputs a ciphertext ${\sf ct}\in\mathcal{C}$, where $\mathcal{C}$ is the cipher space.
		
		\item ${\sf 2SVHSS.Eval}(b,{\sf ek}_b,({\sf ct}^{(1)},\dots,{\sf ct}^{(n)}),f)$: On input a server index $b\in [2]$, an evaluation key ${\sf ek}_b$, a vector of $n$ ciphertexts, a function $f \in {\mathcal{F}}$ with $n$ input values, the homomorphic evaluation algorithm outputs a partial result $y_b$.
		
		\item ${{\sf 2SVHSS.Ver}({\sf vk},(y_1,y_2))}$: On input a verification key $\sf vk$, a pair of partial results $(y_1,y_2)$, the verification algorithm outputs a result $y$ (which is believed to be the correct computation result) or an error symbol $\perp $ (to indicate that one of the servers is cheating).
	\end{itemize}
	
\end{Def}
All parties run the ${\sf 2SVHSS}$ scheme as follows. First, the output client runs ${\sf 2SVHSS.Gen}(1^\lambda)$ to generate $({\sf pk},{\sf vk},({\sf ek}_1,{\sf ek}_2))$. Next, the input clients will run ${\sf 2SVHSS.Enc}({\sf pk},x_i)$ to generate ciphertext ${\sf ct}^{(i)}$ of $x_i$ and upload ${\sf ct}^{(i)}$ to all servers. Then, in order to evaluate a function $f(x_1,\dots,x_n)$, the output client simply sends $f$ to all servers, and the server $b$ runs ${\sf 2SVHSS.Eval}(b,{\sf ek}_b,({\sf ct}^{(i)})_i,f)$ to generate a partial result $y_b$ and returns it to the output client. Finally, the output client runs ${\sf 2SVHSS.Ver}({\sf vk},(y_1,y_2))$ to reconstruct and verify the value of $f(x_1,\dots,x_n)$.

A 2SVHSS scheme should satisfy the following properties: \textit{correctness},  \textit{semantic security}, \textit{verifiablility}, \textit{context hiding} and \textit{compactness}.

\begin{Def}[\bf Correctness]
	The scheme $\sf 2SVHSS$ is said to {\em correctly evaluate} a function family $\mathcal{F}$ if for all honestly generated keys $({\sf pk},{\sf vk},{\sf({ek}_1,{ek_2})})\leftarrow {\sf 2SVHSS.Gen}(1^\lambda)$, for all $x_1,\dots,x_n \in \mathcal{I}$, for all ciphertexts  ${\sf ct}^{(1)},\dots,{\sf ct}^{(n)} \in \mathcal{C}$, where ${\sf ct}^{(i)}\leftarrow{\sf 2SVHSS.Enc}({\sf pk},x_i)$ for $i\in[n]$, for any function $f \in \mathcal{F}$, ${\Pr}_{{\sf 2SVHSS},(x_i)_i,f}^{\sf cor}(\lambda):=\Pr[{\sf 2SVHSS.Ver}({\sf vk},(y_1,y_2))=f(x_1,\dots,x_n)] \geq 1-\lambda^{-\omega(1)},$ where $y_b$ $\leftarrow$ ${\sf 2SVHSS.}$${\sf Eval}(b,{\sf ek}_b,({\sf ct}^{(i)})_i,f)$ for $b\in[2]$ and the probability is taken over all the algorithms’ random choices.
\end{Def}

\begin{Def}[\bf Semantic Security]
	We define the experiment ${\bf Exp}_{\mathcal{A},{\sf 2SVHSS}}^{\sf SS}(1^\lambda)$ with a security parameter $\lambda\in\mathbb{N}$ and a PPT adversary $\mathcal{A}$ as follows:
\begin{flushleft}
		\begin{tabular}{l}
		${\bf Exp}_{\mathcal{A},{\sf 2SVHSS}}^{\sf SS}(1^\lambda):$                         \\
		$(b,x_0,x_1,{\sf state})\leftarrow \mathcal{A}(1^\lambda);\beta\leftarrow\{0,1\}$  \\
		$({\sf pk},{\sf vk},({\sf ek}_1,{\sf ek}_2))\leftarrow{\sf 2SVHSS.Gen}(1^\lambda)$ \\
		${\sf ct}\leftarrow{\sf 2SVHSS.Enc}({\sf pk},x_{\beta})$                           \\
		${\sf input}_b:=({\sf state},{\sf pk},{\sf ek}_b,{\sf ct})$                        \\
		$\beta'\leftarrow\mathcal{A}({\sf input}_b)$                                       \\
		$\text{If } \beta' = \beta \text{ return } 1. \text{ Else return } 0.$
	\end{tabular}
\end{flushleft}
We define the advantage of $\mathcal{A}$ as ${\bf Adv}_{\mathcal{A},{\sf 2SVHSS}}^{\sf SS}(\lambda):=\Pr[{\bf Exp}_{\mathcal{A},{\sf 2SVHSS}}^{\sf SS}(1^\lambda)=1]$. Then we say that $\sf 2SVHSS$ is {\em semantically secure} if for all PPT adversary $\mathcal{A}$ it holds ${\bf Adv}_{\mathcal{A},{\sf 2SVHSS}}^{\sf SS}(\lambda) \leq {\sf negl}(\lambda)$.
\end{Def}

\begin{Def}[\bf Verifiability]
	We define the experiment ${\bf Exp}_{\mathcal{A},{\sf 2SVHSS}}^{\sf Ver}(1^\lambda)$ with a security parameter $\lambda\in\mathbb{N}$ and a PPT adversary $\mathcal{A}$ as follows:
 \begin{itemize}
 	\item ${\sf Setup}$. The challenger runs the $\sf  2SVHSS.Gen(1^\lambda)$ to generate a public key $\sf pk$, a verification key $\sf vk$, a pair of evaluation keys $({\sf ek}_1,{\sf ek}_2)$, and gives $\sf pk$ to $\mathcal{A}$. If $\mathcal{A}$ plays the role of a malicious server $b$ the challenger gives ${\sf ek}_b$ to $\mathcal{A}$.

 	\item ${\sf Verification\,Queries.}$ $\mathcal{A}$ adaptively issues verification queries. Let $(f,(x_i)_i,$ $({\sf ct}^{(i)})_i,y_b')$ be a query from $\mathcal{A}$, where $y_b'$ is a modified partial result and ${\sf ct}^{(i)}\leftarrow{\sf 2SVHSS.Enc}({\sf pk},x_i)$ for all $i\in[n]$. Given the verification query, the challenger proceeds as follows: for each $i\in[n]$ compute $y_{3-b}\leftarrow {\sf 2SVHSS.Eval}(3-b,{\sf ek}_{3-b},({\sf ct}^{(i)})_i,f)$; compute and respond with $y'\leftarrow{\sf 2SVHSS.Ver}({\sf vk},$ $(y'_{b},$ $y_{3-b}))$. In the process of verification queries, if the event $y' \notin \{f(x_1,\dots,$ $x_{n}),\perp\}$ occurs, $\mathcal{A}$ terminates the queries and the experiment outputs 1. If the event never occurs, the experiment outputs 0.
\end{itemize}
	We define the advantage of $\mathcal{A}$ as ${\bf Adv}_{\mathcal{A},{\sf 2SVHSS}}^{\sf Ver}(\lambda):=\Pr[{\bf Exp}_{\mathcal{A},{\sf 2SVHSS}}^{\sf Ver}(1^\lambda)=1]$. We say that $\sf 2SVHSS$ is {\em verifiable under adaptive chosen message and query verification attack}, if for all PPT adversary $\mathcal{A}$ it holds ${\bf Adv}_{\mathcal{A},{\sf 2SVHSS}}^{\sf Ver}(\lambda) \leq {\sf negl}(\lambda)$.
\end{Def}

\begin{Def}[\bf Context Hiding]
	We say that the scheme ${\sf 2SVHSS}$ satisfies {\em context hiding} for a function family $\mathcal{F}$ if there exists a PPT simulator $\sf Sim$ such that the following holds: for any $\lambda \in \mathbb{N}$, any $({\sf pk},{\sf vk},({\sf ek_1},{\sf ek_2}))\leftarrow {\sf 2SVHSS.Gen}(1^\lambda)$, any function $f\in \mathcal{F}$, any  input values $x_1,\dots,x_n\in\mathcal{I}$, any ciphertexts ${\sf ct}^{(1)},\dots,\\{\sf ct}^{(n)}\in\mathcal{C}$, where ${\sf ct}^{(i)}\leftarrow{\sf 2SVHSS.Enc}({\sf pk},x_i)$ for $i\in[n]$, and $y_b \leftarrow {\sf 2SVHSS.Eval}(b,\\{\sf ek}_b, ({\sf ct}^{(i)})_i,$ $f)$ for $b\in[2]$, it holds ${\sf SD}[(y_1,y_2),{\sf Sim}(1^\lambda, {\sf vk}, {\sf pk}, f(x_1,\dots,x_n))]\leq {\sf negl}(\lambda).$
	
\end{Def}

\begin{Def}[\bf Compactness]
	We say that the scheme $\sf 2SVHSS$ {\em compactly evaluates} a function family $\mathcal{F}$ if the running time of ${\sf 2SVHSS.Ver}$ is bounded by a fixed polynomial in $\lambda$.
\end{Def}

\section{A Construction of 2SVHSS}
In this section, we present a construction of 2SVHSS scheme, which allows the clients to outsource the computation of any polynomials $f(x_1,\dots,x_n)$ with ${\sf poly}(\lambda)$ degree, where $\lambda$ is the security parameter. We use the PKE-NLD instantiation in Section 2 to encrypt input values and perform homomorphic computation.

In our construction, an evaluation key consists of two parts: one part is the additive share of the secret key ${\sf sk}$, and the other part is the additive share of the verification key ${\sf vk}=\hat{s}\cdot {\sf sk}$ (where $\hat{s}$ is randomly choose from the secret key distribution). The evaluation algorithm uses the additive share of ${\sf sk}$ to compute the additive share of the output $y$, and uses the additive share of ${\sf vk}$ to compute the additive share of the authentication tag $\tau $ of the output $y$. The verification algorithm uses these additive shares to reconstruct the output $y$ and its tag $\tau$, and then use $\hat{s}$ to verify the output $y$ by checking the equation $\tau =\hat{s}\cdot y$.

 In our construction, the evaluation algorithm consists of 6 subroutines: ${\sf Load}$, ${\sf Add}_1$, ${\sf Add}_2$, ${\sf cMult}$, ${\sf Mult}$, ${\sf Output}$. To compute $f$ the servers need to execute these subroutines ${\sf poly}(\lambda)$ times, and each time these subroutines are executed, there is a unique identifier ${\sf id}\in \mathbb{N}$ corresponding to this execution.

Our scheme ${\sf 2SVHSS}=({\sf 2SVHSS.Gen},{\sf 2SVHSS.Enc},{\sf 2SVHSS.Eval},{\sf 2SVHSS.Ver})$ can be described as follows:

\begin{itemize}

			\item   ${{\sf 2SVHSS.Gen}(1^\lambda)}$: Generate a key pair $({\sf pk},{\sf sk})\leftarrow {\sf PKE.Gen(1^\lambda)}$ for encryption where ${\sf sk}={\bf s}=(1,s)\in R_q^2$. Randomly choose $\hat{s}\leftarrow \mathcal{D}_{\sf sk}$ and let verification key ${\sf vk}=\hat{s}\cdot{\bf s}=(\hat{s},\hat{s}\cdot s)$. Randomly choose ${\bf s}_{1,1}{\leftarrow}R_q^2$ and ${\bf s}_{1,2}{\leftarrow}R_q^2$. Define ${\bf s}_{2,1}={\sf sk}-{\bf s}_{1,1} \mod q,$ ${\bf s}_{2,2}={\sf vk}-{\bf s}_{1,2} \mod q$. Draw two keys $K_1,K_2{\leftarrow} \mathcal{K}^2$ for a pseudorandom function ${\sf PRF}:\mathcal{K}\times\mathbb{N}\rightarrow R_q^2 $. Output ${\sf pk}$, ${\sf vk}$ and $({\sf ek}_1,{\sf ek}_2)$, where ${\sf ek}_b=(K_1,K_2,{\bf s}_{b,1},{\bf s}_{b,2})$ for $b=1,2$.
			
			\item  ${{\sf 2SVHSS.Enc}(1^\lambda,{\sf pk},x)}$: Compute and output ${\bf C}^x\leftarrow {\sf PKE.{\bf OKDM}}({\sf pk},x)$.
			
			\item  	${{\sf 2SVHSS.Eval}(b,{\sf ek}_b,({\bf C}^{x_{1}},\dots,{\bf C}^{x_{n}}),f)}$:
			\begin{itemize}
				\item ${\sf Load}$: On input $({\sf id},{\bf C}^x)$ compute ${\bf t}_b^{x}\leftarrow{\sf PKE}.{\bf DDec}(b,{\bf s}_{b,1},{\bf C}^{x})+(3-2b)\cdot {\sf PRF}(K_1,{\sf id}) \mod q$, ${\bm{\tau}}_b^{x}\leftarrow{\sf PKE}.{\bf DDec}(b,{\bf s}_{b,2},{\bf C}^{x})+(3-2b)\cdot {\sf PRF}(K_2,{\sf id}) \mod q,$ and return ${\bf T}_{b}^{x}=({\bf t}_b^{x},{\bm{\tau}}_b^{x})\in R_q^{2\times 2}$.
				
				\item ${\sf Add_1}$: On input $({\sf id },{\bf T}_b^{x},{\bf T}_b^{x'})$ compute ${\bf t}_b^{x+x'}\leftarrow {\bf t}_b^{x}+{\bf t}_b^{x'}+(3-2b)\cdot {\sf PRF}(K_1,{\sf id}) \mod q$, ${\bm \tau}_b^{x+x'}\leftarrow {\bm \tau}_b^{x}+{\bm \tau}_b^{x'}+(3-2b)\cdot {\sf PRF}(K_2,{\sf id}) \mod q$, and return ${\bf T}_{b}^{x+x'}=({\bf t}_b^{x+x'},{\bm{\tau}}_b^{x+x'})$.
				
				\item ${\sf Add_2}$: On input $({\sf id},{\bf C}^{x},{\bf C}^{x'})$ compute ${\bf C}^{x+x'}\leftarrow {\bf C}^x +{\bf C}^{x'} \mod q$, and return ${\bf C}^{x+x'}$.
				
				\item ${\sf cMult }$: On input $({\sf id},c,{\bf C}^{x})$ compute ${\bf t}_b^{c\cdot x}\leftarrow{\sf PKE}.{\bf DDec}(b,c\cdot{\bf s}_{b,1},{\bf C}^{x})+(3-2b)\cdot {\sf PRF}(K_1,{\sf id}) \mod q$, ${\bm{\tau}}_b^{c\cdot x}\leftarrow{\sf PKE}.{\bf DDec}(b,c\cdot{\bf s}_{b,2},{\bf C}^{x})+(3-2b)\cdot {\sf PRF}(K_2,{\sf id}) \mod q,$ and return ${\bf T}_{b}^{c\cdot x}=({\bf t}_b^{c\cdot x},{\bm{\tau}}_b^{c\cdot x})$.
				
				\item ${\sf Mult }$: On input $({\sf id},{\bf T}_b^{x},{\bf C}^{x'})$ compute ${\bf t}_b^{x\cdot x'}\leftarrow{\sf PKE}.{\bf DDec}(b,{\bf t}_b^{x},{\bf C}^{x'})+(3-2b)\cdot {\sf PRF}(K_1,{\sf id}) \mod q$, ${\bm{\tau}}_b^{x\cdot x'}\leftarrow{\sf PKE}.{\bf DDec}(b,{\bm{\tau}}_b^{x},{\bf C}^{x'})+(3-2b)\cdot {\sf PRF}(K_2,{\sf id}) \mod q,$ and return ${\bf T}_{b}^{x\cdot x'}=({\bf t}_b^{x\cdot x'},{\bm{\tau}}_b^{x\cdot x'})$.
				
				\item ${\sf Output}$: On input $({\sf id},{\bf T}_b^x)$ parses ${\bf T}_b^x=({\bf t}_b^{x},{\bm{\tau}}_b^{x})=((t_{b},\hat{t_b}),(\tau_b,\hat{\tau_b}))$ for some $t_{b},\hat{t_b},\tau_b,\hat{\tau_b}\in R_q$ and output partial result $y_b=(t_b,\tau_b) \mod r.$
			\end{itemize}
			
			\item  	${{\sf 2SVHSS.Ver}({\sf vk},(y_0,y_1))}$: On input verification key ${\sf vk}=(\hat{s},\hat{s}\cdot{ s})$ and two partial results $(y_1,y_2)$, compute $y=t_1+t_2 \mod r$ and $\tau=\tau_1+\tau_2 \mod r$. If $\tau=\hat{s}\cdot y$, output $y$, otherwise, output $\perp$.	
\end{itemize}		
\begin{Theo}
	For all $\lambda\in\mathbb{N}$, for all inputs $x_1,\dots,x_n\in R_r$, for all polynomials $f$ which satisfy: $f$ is of size ${\sf size}(f)\leq {\sf poly}(\lambda)$; the plaintexts upper bound $B_{\sf max}$ with $B_{\sf max}\geq r$, $p/B_{\sf max}\geq \lambda^{\omega(1)}$ and $q/(B_{\sf max}\cdot p)\geq \lambda^{\omega(1)}$; $f$ has maximum number of input addition instructions $P_{{\sf inp}_+}$, for $({\sf pk,vk,(ek_1,ek_2)})\leftarrow{\sf 2SVHSS.Gen}(1^\lambda)$, for ${\bf C}^{x_i}\leftarrow {\sf 2SVHSS.Enc}(1^\lambda,{\sf pk},x_i)$, there exists a PPT adversary $\mathcal{B}$ on the pseudorandom function {\sf PRF} such that ${\Pr}_{{\sf 2SVHSS},(x_i)_i,f}^{\sf cor}(\lambda)\geq1-{\bf Adv}_{{\sf PRF},\mathcal{B}}^{\sf prf}(\lambda)-N\cdot(B_{\sf max}+1)/q-4\cdot{\sf size}(f) \cdot N^2\cdot P_{\sf inp+}\cdot B_{\sf max}\cdot(B_{\sf ct}\cdot p/q+B_{\sf sk}^2/p)-4\cdot{\sf size}(f)\cdot N\cdot (p/q+1/p)$.
\end{Theo}
\begin{proof}
	First, let $\epsilon_0=\Pr_{{\sf 2SVHSS},(x_i)_i,f}^{\sf cor}(\lambda)$. Our goal is to prove that for all inputs $x_1,\dots,x_n\in{R_r}$ and for all polynomials $f$, the probability $|1-\epsilon_0|\leq{\sf negl}(\lambda)$. And, let $\epsilon_1=\Pr_{{\sf 2SVHSS},(x_i)_i,f}^1(\lambda)$ denote the probability that evaluation yields the correct output, where we replace every evaluation of the PRF by inserting a value ${\bf r}\leftarrow R_{q}^2$ chosen at random. Boyle et al. \cite{BKS19} proved that $|\epsilon_0-\epsilon_1|\leq{\bf Adv}_{\sf PRF,\mathcal{B}}^{\sf prf}(\lambda).$
	
	Next, we give a lower bound of the probability $\epsilon_1$. It is for this reason that we prove that with overwhelming probability over the choice of ${\bf r}\leftarrow R_q^2$ all shares $({\bf T}_{1}^{x},{\bf T}_{2}^{x})$ computed during homomorphic evaluation of $f$ satisfy ${\bf t}_1^x+{\bf t}_2^x=x\cdot {\bf s}=(x,x\cdot s) \mod q$ (1), ${\bm \tau}_1^x+{\bm \tau}_2^x=x\cdot\hat{s} \cdot {\bf s}=(x\cdot \hat{s},x\cdot\hat{s}\cdot s) \mod q$ (2). For $m\in R$ and $z_1,z_2 \in R_q$ be random, $z_1+z_2=m$ over $R_r$ with probability at least $1-N\cdot(B_{\sf max}+1)/q\geq 1-\lambda^{-\omega(1)}$, which has proved in \cite{BKS19}. Therefore assuming (1) and (2) are true, $t_1+t_2=x $ and $\tau_1+\tau_2=x\cdot \hat{s}$ over $R_r$ with probability at least $1-N\cdot(B_{\sf max}+1)/q$. It is left to prove that indeed (1) and (2) hold true during homomorphic evaluation of $f$. ${\sf PKE}.{\bf DDec}$ is the procedure for distributed decryption. Under the assumption that distributed decryption is always successful, we prove that the subroutines of evaluation algorithm and verification algorithm preserves correctness. Because addition of input values and the output of a memory value does not affect share, we ignore them.
\begin{itemize}
	\item  Consider input $({\sf id},{\bf C}^x)$ for $b\in[2]$. We have ${\bf t}_1^x+{\bf t}_2^x={\sf PKE}.{\bf DDec}(1,{\bf s}_{1,1},{\bf C}^x)+{\bf r}+{\sf PKE}.{\bf DDec}(2,{\bf s}_{2,1},{\bf C}^x)$ $-$ ${\bf r}\mod q=x\cdot {\bf s} \mod q$, ${\bm{\tau}}_1^x+{\bm \tau}_2^x={\sf PKE}.$ ${\bf DDec}(1,{\bf s}_{1,2},{\bf C}^x)$ $+$ ${\bf r}+{\sf PKE}.{\bf DDec}(2,{\bf s}_{2,2},{\bf C}^x)-{\bf r}\mod q =x\cdot\hat{s}\cdot{\bf s} \mod q$.
	\item  Consider input $({\sf id },{\bf T}_b^{x},{\bf T}_b^{x'})$ for $b\in[2]$. We have ${\bf t}_1^{x+x'}+{\bf t}_2^{x+x'}={\bf t}_1^{x}+{\bf t}_1^{x'}+{\bf r}+{\bf t}_2^{x}+{\bf t}_2^{x'}-{\bf r} \mod q	=x\cdot {\bf s}+x'\cdot {\bf s} \mod q =(x+x')\cdot {\bf s} \mod q$, ${\bm \tau}_1^{x+x'}+{\bm \tau}_2^{x+x'}={\bm \tau}_1^{x}+{\bm \tau}_1^{x'}+{\bf r}+{\bm \tau}_2^{x}+{\bm \tau}_2^{x'}-{\bf r} \mod q =x\cdot\hat{s}\cdot {\bf s}+x'\cdot\hat{s}\cdot {\bf s} \mod q =(x+x')\cdot\hat{s}\cdot {\bf s} \mod q.$
	\item  Consider input  $({\sf id},c,{\bf C}^{x})$ for $b\in[2]$. We have ${\bf t}_1^{c\cdot {x}}+{\bf t}_2^{c\cdot {x}}={\sf PKE}.{\bf DDec}(1,c\cdot{\bf s}_{1,1},{\bf C}^{x})+{\bf r}+{\sf PKE}.{\bf DDec}(2,c\cdot{\bf s}_{2,1},{\bf C}^{x})-{\bf r}\mod q=(c\cdot{\bf s})\cdot x \mod q=(c\cdot x)\cdot {\bf s} \mod q$, ${\bm \tau}_1^{c\cdot {x}}+{\bm \tau}_2^{c\cdot {x}}={\sf PKE}.{\bf DDec}(1,c\cdot{\bf s}_{1,2},{\bf C}^{x})+{\bf r}+{\sf PKE}.{\bf DDec}(2,c\cdot{\bf s}_{2,2},{\bf C}^{x})-{\bf r}\mod q=(c\cdot\hat{s}\cdot{\bf s})\cdot x \mod q =(c\cdot x)\cdot\hat{s}\cdot {\bf s} \mod q$.
	\item  Consider input  $({\sf id},{\bf T}_b^{x},{\bf C}^{x'})$ for $b\in[2]$. Assuming correctness holds for shares $({\bf T}_{1}^{x},{\bf T}_{2}^{x})$ and distributed decryption it holds ${\bf t}_1^{x\cdot {x'}}+{\bf t}_2^{x\cdot {x'}}={\sf PKE}.{\bf DDec}(1,{\bf t}_{1}^x,$ ${\bf C}^{x'})+{\bf r}+{\sf PKE}.{\bf DDec}(2,{\bf t}_{2},{\bf C}^{x'})-{\bf r}\mod q=(x\cdot {\bf s})\cdot x' \mod q =(x\cdot x')\cdot {\bf s} \mod q$, ${\bm \tau}_1^{x\cdot {x'}}+{\bm \tau}_2^{x\cdot {x'}}={\sf PKE}.{\bf DDec}(1,{\bm \tau}_{1}^x,{\bf C}^{x'})+{\bf r}+{\sf PKE}.{\bf DDec}(2,{\bm \tau}_{2},{\bf C}^{x'})-{\bf r}\mod q=(x\cdot\hat{s}\cdot{\bf s})\cdot x' \mod q =(x\cdot x')\cdot\hat{s}\cdot {\bf s} \mod q$.
\end{itemize}
From above all, for verification algorithm output $y_b=(t_b,\tau_b)\mod r$ ($b\in[2]$) it holds $\tau=\tau_{1}+\tau_{2}=\hat{s}\cdot t_{1}+\hat{s}\cdot t_{2}=\hat{s}\cdot(t_1+t_2)=\hat{s}\cdot y$. As the equality $\tau=\hat{s}\cdot y$ is always satisfied, the verification algorithm will output $y=f(x_1,\dots, x_n)$ with probability 1.

	Last, we need to bound the probability that distributed decryption fails. By Section 2.1, the distributed decryption fails with probability at most
	$N^2\cdot P_{\sf inp+}\cdot\|x\|_\infty\cdot B_{\sf ct}\cdot p/q+N\cdot \|x\cdot m\|_\infty/p+N\cdot(p/q+1/p).$
	Throughout the evaluation of $f$ we are guaranteed $\|x\|\leq B_{\sf max}$ for all intermediary values $x\in R$. We need to give the upper bound of $\|x\cdot m\|_\infty$. For the messages $m_i=x_i\cdot s_{j_i}$ we have
	$\|x\cdot \sum_{i=1}^{P_{\sf inp+}}x_i\cdot s_{j_i}\|_{\infty}\leq\sum_{i=1}^{P_{\sf inp+}}\|x\cdot x_i\cdot s_{j_i}\|_{\infty}\leq P_{\sf inp+}\cdot N\cdot B_{\sf max}\cdot B_{\sf sk}.$
	For the messages $m_i=x_i\cdot \hat{s} \cdot s_{j_i}$ we have
	$\|x\cdot \sum_{i=1}^{P_{\sf inp+}}x_i\cdot \hat{s} \cdot s_{j_i}\|_{\infty}\leq\sum_{i=1}^{P_{\sf inp+}}\|x\cdot x_i\cdot \hat{s}\cdot s_{j_i}\|_{\infty}\leq P_{\sf inp+}\cdot N\cdot B_{\sf max}\cdot B_{\sf sk}^2.$
	
	Finally, applying a union bound over all $4\cdot{\sf size}(f)$ decryptions (one homomorphic multiplication corresponds to 4 decryptions) yields
	$\epsilon_1\geq 1-N\cdot(B_{\sf max}+1)/q-4\cdot{\sf size}(f) \cdot N^2\cdot P_{\sf inp+}\cdot B_{\sf max}\cdot(B_{\sf ct}\cdot p/q+B_{\sf sk}^2/p)-4\cdot{\sf size}(f) \cdot N\cdot (p/q+1/p)$.\qed

\end{proof}

\begin{Theo}
	The scheme $\sf 2SVHSS$ is semantically secure. 	
\end{Theo}
The proof for the semantic security of our 2SVHSS is quite similar to that of [9]. In order to avoid duplication we only provide a proof sketch.

\begin{proofsketch}
We prove that for every PPT adversary $\mathcal{A}$ on the semantic security of $\sf 2SVHSS$ there exists a PPT adversary $\mathcal{B}$ on the security of $\sf PKE.OKDM$ such that ${\bf Adv}_{\mathcal{A},\sf 2SVHSS}^{\sf ss}(\lambda)\leq {\bf Adv}_{{\mathcal{B},\sf PKE.OKDM}}^{\sf kdm-ind}(\lambda)$. Boyle et al.\cite{BKS19} have proved ${\bf Adv}_{\mathcal{A},\sf HSS}^{\sf ss}(\lambda)\leq {\bf Adv}_{\mathcal{B},{\sf PKE.OKDM}}^{\sf kdm-ind}(\lambda)$. Next we will explain that ${\bf Adv}_{\mathcal{A},\sf 2SVHSS}^{\sf ss}(\lambda)={\bf Adv}_{\mathcal{A},\sf HSS}^{\sf ss}(\lambda)$. The ciphertexts of {\sf 2SVHSS} are generated in the same way as {\sf HSS} of \cite{BKS19}. But the information obtained by the adversary $\mathcal{A}$ in {\sf 2SVHSS} is different from {\sf HSS}, because the evaluation key in {\sf 2SVHSS} is not exactly the same as the evaluation key in {\sf HSS}, more specifically, the latter has more random numbers than the former. But the extra random numbers in the evaluation key of {\sf 2SVHSS} does not provide any additional information about the input value to the adversary $\mathcal{A}$, therefore we have that ${\bf Adv}_{\mathcal{A},\sf 2SVHSS}^{\sf ss}(\lambda)={\bf Adv}_{\mathcal{A},\sf HSS}^{\sf ss}(\lambda)\leq {\bf Adv}_{{\mathcal{B},\sf PKE.OKDM}}^{\sf kdm-ind}(\lambda).$ \qed
\end{proofsketch}

\begin{Theo}
	The scheme $\sf 2SVHSS$ is verifiable.
\end{Theo}
\begin{proof}
	Let $A$ be the event that ${\bf Exp}_{\mathcal{A},{\sf 2SVHSS}}^{\sf Ver}(1^\lambda)$ outputs 1.  Let $A_j$ be the event that the ${\bf Exp}_{\mathcal{A},{\sf 2SVHSS}}^{\sf Ver}(1^\lambda)$ outputs 1 after $j$ verification queries. Let $Q$ be the upper bound on the number of verification queries requested by the adversary. To prove the theorem we only need to prove $\Pr[A]\leq{\sf negl}(\lambda)$.
	
	We first give the probability of event $A_j$ happening. In the $j$-th verification query, the adversary $\mathcal{A}$ sends  $(f,(x_i)_i,({\sf ct}^{(i)})_i,y_b')$ (where $b\in[2]$, if $\mathcal{A}$ plays the role of a malicious first server $b=1$, otherwise $b=2$) to the challenger, where $y_b'=(t_b',\tau_b')$. The challenger computes $y_{3-b}\leftarrow {\sf 2SVHSS.Eval}(3-b,{\sf ek}_{3-b},({\sf ct}^{(i)})_i,f)$ for all $i\in[n]$, where $y_{3-b}=(t_{3-b},\tau_{3-b})$. Next, the challenger runs ${\sf 2SVHSS.Ver}({\sf vk},(y_1,y_2))$ to compute $y'=t_b'+t_{3-b}$ and $\tau'=\tau_b'+\tau_{3-b}$.
	
	Denote $y$ and $\tau$ as the correct value when all participants honestly evaluate $f$, and let $\Delta_y=y'-y$ and $\Delta_\tau=\tau'-\tau$. Then the event $A_i$ occurs if $\Delta_y \neq 0$ and $\tau'=\hat{s}\cdot y'$, that is, $\hat{s}\cdot\Delta_y=\Delta_\tau$. $\mathcal{A}$ can choose $\Delta_y$ and $\Delta_\tau$ by modifying $y_b'=(t_b',\tau_b')$. Hence, the only way for $\mathcal{A}$ to let the challenger accept a wrong result is to guess $\hat{s}$. Since $\hat{s}$ is a polynomial of degree $N$ and its coefficients are uniform in $\{0, \pm1\}$, subject to the constraint that only $h_{\sf sk}=N/2$ coefficients are non-zero, then there are $P=2^{h_{\sf sk}}\binom{N}{h_{\sf sk}}=2^{h_{\sf sk}}\frac{\prod_{i=0}^{h_{\sf sk}-1}(N-i)}{h_{\sf sk}!}=2^{h_{\sf sk}}\prod_{i=0}^{h_{\sf sk}-1}\frac{N-i}{h_{\sf sk}-i}\geq2^{h_{\sf sk}}\prod_{i=0}^{h_{\sf sk}-1}\frac{N}{h_{\sf sk}}=2^N$ possible values for $\hat{s}$. After $j-1$ queries $\mathcal{A}$ can exclude $j-1$ impossible values of $\hat{s}$, which means the number of possible values for $\hat{s}$ is $P-(j-1)$. So $\Pr[A_i]=\frac{1}{P-(j-1)}=\frac{1}{P-j+1}$.
	
	From above all, we have $\Pr[A]=\Pr[\bigcup_{j=1}^Q A_i] \leq \sum_{j=1}^Q \Pr[A_i]=\sum_{j=1}^Q\frac{1}{P-j+1}$ $\leq\sum_{j=1}^Q \frac{1}{P-Q+1}\leq\frac{Q}{P-Q}$. Because $N>\lambda$ when choosing parameters (see Table 2 for more information), $P\geq2^{N}>2^\lambda$, and $Q={\sf poly}(\lambda)$, $\frac{Q}{P-Q}\leq{\sf negl}(\lambda)$. \qed
\end{proof}

\begin{Theo}
	The scheme $\sf 2SVHSS$ satisfies context hiding.
\end{Theo}
\begin{proof}
	We show how to construct a simulator $\sf Sim$ that can generate $(y_1',y_2')$ with negligible statistical distance from $(y_1,y_2)$, where $y_b=(t_b,\tau_b)$ for $b\in[2]$.
	
	We give the description of simulator ${\sf Sim}$: on input the security parameter $1^\lambda$, the verification key ${\sf vk}=(\hat{s},\hat{s}\cdot s)$, the public key $\sf pk$ and the $y=f(x_1,\dots,x_n)$, the simulator ${\sf Sim}$ chooses $t_1'{\leftarrow} R_q$ and $\tau_1'{\leftarrow} R_q$ at random, and let $t_2'=y-t_1' \mod q$, $\tau_2'=\hat{s}\cdot y-\tau_1' \mod q$. The simulator $\sf Sim$ outputs $(y_1',y_2')$, where $y_b'=(t_b',\tau_b')$ for $b\in[2]$. It is straightforward to see that  $(y_1,y_2)$ is indistinguishable from the $(y_1',y_2')$.\qed
\end{proof}

\section{Performance Analysis}
In this section, we implemented our scheme and got the running time of our scheme, and compared our scheme with the LMS scheme \cite{LMS18} in terms of efficiency.
\subsection{Evaluating {2SVHSS}}
We have implemented the scheme ${\sf 2SVHSS}$  in a Ubuntu 18.04.2LTS 64-bit operating system with ${\rm Intel}^{\circledR}$ Xeon(R) Gold 5218 2.30GHZ$\times$64 processors and 160GB RAM. We choose the PRF as the standard AES with 128-bit secret key from the library OpenSSL 1.0.2g and realize all large integer related mathematical computations based on the C libraries GMP 6.1.2 and FLINT 2.5.2.

According to the method of selecting parameters provided by \cite{BKS19}, we first choose the plaintexts upper bound $B_{\sf max}$ and let $r=B_{\sf max}$. $r$ can take any integer in $[2,B_{\sf max}]$, in order to maximize the input space $\mathcal{I}=R_r$ we let $r=B_{\sf max}$. Next we choose $B_{\sf err}=8\sigma$, $\sigma=8$ for the noise distribution, a statistical security parameter $\kappa=40$, the number non-zero entries in the secret key $h_{\sf sk}=N/2$, the number of homomorphic additions of inputs $P_{\sf inp+}=1$, and $B_{\sf sk}=1$. We choose the parameters so that each multiplication has a failure probability no more than $2^{-\kappa}$. To ensure this holds we set $p=N\cdot B_{\sf max}\cdot h_{\sf sk}\cdot 2^{\kappa+2}$ and $q\geq 2^{-\kappa+3}\cdot p \cdot N^2 \cdot B_{\sf max}\cdot B_{\sf ct}$, where $B_{\sf ct}=B_{\sf err}(2h_{\sf sk}+1)$ proved by \cite{BKS19}. The security parameters are obtained through the LWE estimator tool\cite{LWEtool} by Albrecht et al.\cite{APS15}. The parameters corresponding to different $B_{\sf max}$ and the running time of 6 subroutines in {\sf 2SVHSS.Eval} algorithm are listed in Table 2.

\begin{table}[]
	\centering
	\caption{The Running Time(in milliseconds) of 6 Subroutines in {\sf 2SVHSS.Eval}}
	\begin{tabular}{ccccccccccc}
		\hline
		$B_{\sf max}$ & $N$     & $\lg p$ & $\lg q$ & Security & $\sf Load$ & $\sf Add_1$ & $\sf Add_2$  & $\sf cMult$& $\sf Mult$ & $\sf Output$\\ \hline
		2       & 4096  & 66   & 153  & 117.1    & 106      & 13  & $<1$  & 105   & 105  & $<1$   \\
		$2^{16}$     & 4096  & 81   & 183  & 86.5     & 114      & 13  & $<1$  & 115   & 114  & $<1$   \\
		$2^{32}$      & 8192  & 99   & 220  & 198.7    & 276      & 29   & $<1$ & 275   & 274  & $<1$   \\
		$2^{64}$      & 8192  & 131  & 284  & 128.9    & 315      & 29  & $<1$   & 320  & 318  & $<1$   \\
		$2^{128}$     & 16384 & 197  & 417  & 214.0    & 1623     & 67  & $<1$  & 1633  & 1630  & $<1$  \\
		$2^{256}$     & 16384 & 325  & 673  & 96.7     & 2049     & 69   & $<1$  & 2016  & 2012 & $<1$   \\ \hline
	\end{tabular}
\end{table}

\subsection{Comparisons with LMS\cite{LMS18}}
We compare the efficiency of the scheme ${\sf2SVHSS}$ with the LMS scheme \cite{LMS18} which supports polynomials of highest degree among all existing works, when the same number of servers are used. Because LMS is based on the $k$-HE assumption, we choose the homomorphic encryption scheme ${\sf SH}$ \cite{BV11} to implement LMS. The reason for choosing the scheme ${\sf SH}$ is to achieve a fair comparison, because the scheme $\sf PKE$ in ${\sf2SVHSS}$ is also based on ${\sf SH}$. Under different $B_{\sf max}$ and the degree of the polynomials to be computed, the parameters $N$, $q$ and security parameters of LMS are different. We set those parameters according to the method of \cite{NLV11}, and use the LWE estimator tool\cite{LWEtool,APS15} to estimate the corresponding security parameters. The parameters of ${\sf2SVHSS}$ and LMS \cite{LMS18} are shown in Table 3. It is fair to compare the two schemes under the same security parameters. But in the case of using LWE estimator tool to estimate the safety parameters, it is difficult to ensure that the safety parameters are exactly the same. To avoid the suspicion of deliberately exaggerating the efficiency of ${\sf 2SVHSS}$ we let the security parameters of ${\sf 2SVHSS}$ larger than those of LMS under the same $B_{\sf max}$.

\begin{table}[]
	\centering
	\caption{The Parameters of LMS and $\sf 2SVHSS$}
		\begin{tabular}{ccccccccccc}
			\hline
			$B_{\sf max}$             &              & $N$  & lg $q$ & Security & \qquad\quad & $B_{\sf max}$              &              & $N$   & lg $q$ & Security \\ \hline
			\multirow{5}{*}{$2^{32}$} & LMS deg-5    & 4096 & 151    & 119.5    & \qquad\quad & \multirow{5}{*}{$2^{128}$} & LMS deg-5    & 8192  & 441    & 62.8     \\
			& LMS deg-7    & 4096 & 202    & 73.5     & \qquad\quad &                            & LMS deg-7    & 8192  & 589    & 42.3     \\
			& LMS deg-9    & 8192 & 258    & 151.9    & \qquad\quad &                            & LMS deg-9    & 16384 & 742    & 82.6     \\
			& LMS deg-11   & 8192 & 311    & 111      & \qquad\quad &                            & LMS deg-11   & 16384 & 892    & 62       \\
			& {\sf 2SVHSS} & 8192 & 220    & 198.7    & \qquad\quad &                            & {\sf 2SVHSS} & 16384 & 417    & 214.0    \\ \hline
			\multirow{5}{*}{$2^{64}$} & LMS deg-5    & 4096 & 246    & 53.6     & \qquad\quad & \multirow{5}{*}{$2^{256}$} & LMS deg-5    & 16384 & 827    & 69.5     \\
			& LMS deg-7    & 8192 & 334    & 99       & \qquad\quad &                            & LMS deg-7    & 16384 & 1104   & 46.1     \\
			& LMS deg-9    & 8192 & 417    & 69.6     & \qquad\quad &                            & LMS deg-9    & 32768 & 1387   & 92.2     \\
			& LMS deg-11   & 8192 & 502    & 51.5     & \qquad\quad &                            & LMS deg-11   & 32768 & 1665   & 69.5     \\
			& {\sf 2SVHSS} & 8192 & 284    & 128.9    & \qquad\quad &                            & {\sf 2SVHSS} & 16384 & 673    & 96.7     \\ \hline
		\end{tabular}
\end{table}

\hspace{-0.5cm}{\bf Theoretical Analysis.} On the server-side, to compute a degree-$d$ term, LMS needs two servers to compute $2^{d-1}$ degree-$d$ terms respectively, which will cause LMS to be very slow when computing high degree polynomials. $\sf 2SVHSS$ does not have this problem. On the client-side, for LMS, as the polynomial degree increases, the ciphertext size will also become larger (Table 3 shows this intuitively), which makes the time spent by LMS in decryption will increase as the degree of polynomials increases. $\sf 2SVHSS$ does not have the problem of ciphertext size increase, so the time spent by $\sf 2SVHSS$ is fixed.

\begin{table}[]
	\centering
	\caption{The Running Time (in seconds) of LMS and $\sf 2SVHSS$}
	\begin{tabular}{ccccccccc}
		&&&&Server-Side \\
		\hline
		The degree           & \multicolumn{2}{c}{deg-5}                   & \multicolumn{2}{c}{deg-7}                   & \multicolumn{2}{c}{deg-9}                   & \multicolumn{2}{c}{deg-11}                  \\ \hline
		{$B_{\sf max}$}& LMS                  & {\sf 2SVHSS}         & LMS        & {\sf 2SVHSS}         & LMS                  & {\sf 2SVHSS}         & LMS                  & {\sf 2SVHSS}         \\
		\hline
		$2^{32}$             & 0.458                & 1.145                & 6.230                & 1.710                & 124.592              & 2.295                & 1016.660             & 2.831                \\
		$2^{64}$             & 0.588                & 1.152                & 19.230               & 1.849                & 159.609              & 2.700                & 1618.510             & 3.256                \\
		$2^{128}$            & 2.288                & 2.931                & 58.317               & 4.559                & 874.877              & 5.683                & 7620.583             & 6.905                \\
		$2^{256}$            & 14.941               & 10.761               & 214.354              & 14.392               & 3787.569             & 18.532               & 31097.442            & 22.527               \\ \hline
		&&&&Client-Side \\
		\hline
		$2^{256}$            & 0.129               & 0.0024               & 0.427              & 0.0024              & 1.497             & 0.0024               & 2.134            & 0.0024\\
		\hline
		\multicolumn{1}{l}{} & \multicolumn{1}{l}{} & \multicolumn{1}{l}{} & \multicolumn{1}{l}{} & \multicolumn{1}{l}{} & \multicolumn{1}{l}{} & \multicolumn{1}{l}{} & \multicolumn{1}{l}{} & \multicolumn{1}{l}{} \\
		\multicolumn{1}{l}{} & \multicolumn{1}{l}{} & \multicolumn{1}{l}{} & \multicolumn{1}{l}{} & \multicolumn{1}{l}{} & \multicolumn{1}{l}{} & \multicolumn{1}{l}{} & \multicolumn{1}{l}{} & \multicolumn{1}{l}{}\\	
	\end{tabular}
\vspace{-1cm}
\end{table}

\hspace{-0.5cm}{\bf Experimental Results.} Table 4 shows the server-side and client-side running time of LMS and ${\sf 2SVHSS}$ computing one degree-$d$ term for $d\in\{5,7,9,11\}$, where server-side time is the average running time of the two servers to execute the evaluation algorithm.

It is easy to find out from Table 4 that when computing low-degree polynomials with a small $B_{\sf max}$, LMS has trivial advantage on server-side. But ${\sf 2SVHSS}$ has a enormous advantage when computing polynomials higher than degree-7. A large amount of server-side running time of LMS causes the client to wait a long time to obtain the result, which makes LMS difficult to apply in practice.

The advantage of ${\sf 2SVHSS}$ on the client-side is obvious. The cost of computing the degree-5 term LMS scheme is about 54 times that of ${\sf 2SVHSS}$. And as the degree of polynomial increases, this advantage becomes more obvious, which consistent with our theoretical analysis.

\section{Concluding Remarks}
In order to solve the problem that the clients cannot compute and verify high-degree polynomial functions over outsourced data on a small number of servers, we proposed a two-server verifiable secret sharing model and constructed a scheme in this model. Our scheme  allows the clients to efficient compute and verify the value of polynomials that may have a degree as high as a polynomial in the security parameter. In addition, $\sf 2SVHSS$ can protect outsourced data from leaking to the servers and the output client. In practical applications, $\sf 2SVHSS$ is better than the current best scheme LMS in computing high-degree polynomial functions.

\hspace{-0.52cm}{\bf Acknowledgments.} The research was supported by Singapore Ministry of Education under Research Grant RG12/19 and National Natural Science Foundation of China (No. 61602304).

\end{document}